\documentclass[12pt]{article}

\begin{document}

\title{\bf A Classification of Plane Symmetric Kinematic
Self-Similar Solutions}

\author{M. Sharif \thanks{msharif@math.pu.edu.pk} and Sehar Aziz
\thanks{sehar$\_$aziz@yahoo.com}\\
Department of Mathematics, University of the Punjab,\\
Quaid-e-Azam Campus, Lahore-54590, Pakistan.}

\date{}

\maketitle

\begin{abstract}
In this paper we provide a classification of plane symmetric
kinematic self-similar perfect fluid and dust solutions. In the
perfect fluid and dust cases, kinematic self-similar vectors for
the tilted, orthogonal and parallel cases have been explored in
the first, second, zeroth and infinite kinds with different
equations of state. We obtain total of eleven plane symmetric
kinematic self-similar solutions out of which six are independent.
The perfect fluid case gives two solutions for infinite tilted and
infinite orthogonal kinds. In the dust case, we have four
independent solutions in the first orthogonal, infinite tilted,
infinite orthogonal and infinite parallel kinds. The remaining
cases provide contradiction. It is interesting to mention that
some of these solutions turn out to be vacuum.
\end{abstract}

{\bf Keywords:} Plane symmetry, Self-similar variable.

\section{Introduction}

General Theory of Relativity (GR), which is a field theory of
gravitation and is described in terms of geometry, is highly
non-linear. Because of this non-linearity, it becomes very
difficult to solve the gravitational field equations unless
certain symmetry restrictions are imposed on the spacetime metric.
These symmetry restrictions are expressed in terms of isometries
possessed by spacetimes. These isometries, which are also called
Killing Vectors (KVs), give rise to conservation laws [1].

There has been a recent literature [2-8, and references therein]
which shows a significant interest in the study of various
symmetries. These symmetries arise in the exact solutions of
Einstein field equations (EFEs) given by
\begin{equation}
G_{ab}=R_{ab}-\frac 12 Rg_{ab}=\kappa T_{ab},
\end{equation}
where $G_{ab}$ represents the components of Einstein tensor,
$R_{ab}$ are the components of Ricci tensor and $T_{ab}$ are the
components of matter (or energy-momentum) tensor, $R$ is the Ricci
scalar and $\kappa$ is the gravitational constant. The geometrical
nature of a spacetime is expressed by the metric tensor through
EFEs.

Self-similarity is very helpful in simplifying the field
equations. Self-similarity leads to the reduction of the governing
equations from partial differential equations to ordinary
differential equations, whose mathematical treatment is relatively
simple. Although self-similar solutions are only special
solutions, they often play an important role as an intermediate
attractor. There does not exist any characteristic scale in
Newtonian gravity or GR. Invariance of the field equations under a
scale transformation indicates that there exist scale invariant
solutions to the EFEs. These solutions are known as self-similar
solutions.

In order to obtain realistic solutions of gravitational collapse
leading to star formation, self-similar solutions have been
investigated by many authors in Newtonian gravity [9]. There exist
several preferred geometric structures in self-similar models, a
number of natural approaches may be used in studying them. The
three most common ones are the co-moving, homothetic and
Schwarzschild approaches. In this paper, we shall use the
co-moving approach. In the co-moving approach, pioneered by Cahill
and Taub [10], the coordinates are adopted to the fluid 4-velocity
vector. This probably affords the best physical insight and is the
most convenient one. In GR, self-similarity is defined by the
existence of a homothetic vector (HV) field. Such similarity is
called the first kind (or homothety or continuous self similarity
(CSS)). There exists a natural generalization of homothety called
kinematic self-similarity, which is defined by the existence of a
kinematic self-similar (KSS) vector field. Kinematic
self-similarity is characterized by an index $\alpha/\delta$
(similarity index) and can be classified into three kinds. The
basic condition characterizing a manifold vector field $\xi$ as a
self-similar generator is given by
\begin{eqnarray}
\pounds_\xi A=\lambda A,
\end{eqnarray}
where $\lambda$ is a constant, $A$ is independent physical field
and $\pounds_{\xi}$ denotes the Lie derivative along $\xi$. This
field can be scalar (e.g., $\mu$), vector (e.g., $u_{a}$) or
tensor (e.g., $g_{ab}$). In GR, the gravitational field is
represented by the metric tensor $g_{ab}$, and an appropriate
definition of geometrical self-similarity is necessary.

The vector field $\xi$ can have three cases, i.e., parallel,
orthogonal and tilted. They are distinguished by the relation
between the generator and a timelike vector field, which is
identified as the fluid flow, if it exists. The tilted case is the
most general among them.

The self-similar idea of Cahil and Taub [10] corresponds to
Newtonian self-similarity of the homothetic class. Carter and
Henriksen [11,12] defined the other kinds of self-similarity
namely second, zeroth and infinite kinds. In the context of
kinematic self-similarity, homothety is considered as the first
kind. Several authors have explored KSS perfect fluid solutions.
The only compatible barotropic equation of state with
self-similarity of the first kind is
\begin{eqnarray}
p=k\rho.
\end{eqnarray}
Carr [2] has classified the self-similar perfect fluid solutions
of the first kind in the dust case ($k=0$). The case $0<k<1$ has
been studied by Carr and Coley [3]. Coley [13] has shown that the
FRW solution is the only spherically symmetric homothetic perfect
fluid solution in the parallel case. McIntosh [14] has discussed
that a stiff fluid ($k=1$) is the only compatible perfect fluid
with the homothety in the orthogonal case. Benoit and Coley [15]
have studied analytic spherically symmetric solutions of the EFE's
coupled with a perfect fluid and admitting a KSS vector of the
first, second and zeroth kinds.

Carr et al. [16] have considered the KSS associated with the
critical behavior observed in the gravitational collapse of
spherically symmetric perfect fluid with equation of state
$p=k\rho$. They showed for the first time the global nature of
these solutions and showed that it is sensitive to the value of
$\alpha$. Carr et al. [17], further, investigated solution space
of self-similar spherically symmetric perfect fluid models and
physical aspects of these solutions. They combine the state space
description of the homothetic approach with the use of the
physically interesting quantities arising in the co-moving
approach. Coley and Goliath [18] have investigated self-similar
spherically symmetric cosmological models with a perfect fluid and
a scalar field with an exponential potential.

Gravitational collapse is one of the fundamental problems in GR.
Self-similar gravitational collapse and critical collapse provides
information about the collapse of an object. The collapse
generally has three kinds of possible final states. First is
simply the halt of the processes in a self-sustained object or the
description of a matter field or gravitational field. The second
is the formation of black holes with outgoing gravitational
radiation and matter, while the third is the formation of naked
singularities. Critical collapse in the context of self-similarity
gives the information about the mass of black holes formed as a
result of collapse.

Recently, Maeda et al. [4,5] studied the KSS vector of the second,
zeroth and infinite kinds in the tilted case. They assumed the
perfect fluid spacetime obeying a relativistic polytropic equation
of state. Further, they assumed two kinds of polytropic equation
of state and showed that such spacetimes must be vacuum in both
cases. They explored the case in which a KSS vector is not only
tilted to the fluid flow but also parallel or orthogonal. The same
authors [6] have also discussed the classification of the
spherically symmetric KSS perfect fluid and dust solutions. This
analysis has provided some interesting solutions.

In a recent paper, Sharif and Sehar [8] have investigated the KSS
solutions for the cylindrically symmetric spacetimes. The analysis
has been extensively given for the perfect and dust cases with
tilted, parallel and orthogonal vectors by using different
equations of state. Some interesting consequences have been
developed. The same authors have also studied the properties of
such solutions for spherically symmetric [19], cylindrically
symmetric [20] and plane symmetric spacetimes [21].

The group $G_3$ contains two special cases of particular physical
interest: spherical and plane symmetry. In this paper we shall use
the same procedure to investigate KSS solutions for the plane
symmetric spacetimes. The paper has been organised as follows. In
section 2, we shall discuss KSS vector of different kinds for the
plane symmetric spacetimes. Section 3 is devoted to titled perfect
fluid case. In section 4, we shall find out the titled dust
solutions. Sections 5 and 6 are used to explore the orthogonal
perfect fluid and dust solutions respectively. Sections 7 and 8
are devoted to parallel perfect fluid and dust cases. Finally, we
shall summarise and discuss all the results.

\section{Plane Symmetric Spacetime and Kinematic Self-Similarity}

A plane symmetric Lorentzian manifold is defined to be the
manifold which admits the group $SO(2)\times \Re^2$  as the
minimal isometry group in such a way that the group orbits are
spacelike surfaces of zero curvature, where $SO(2)$ corresponds to
a rotation and $\Re^2$ to the translations along spatial
directions $y$ and $z$. The metric for the most general plane
symmetric spacetime has the following form [22]
\begin{equation}
ds^2=e^{2\nu(t,x)}dt^2-e^{2\mu(t,x)}dx^2-e^{2\lambda(t,x)}(dy^2+
dz^2),
\end{equation}
where $\nu$, $\mu$ and $\lambda$ are arbitrary functions of $t$
and $x$. It has three isometries given as
$\xi_1=\partial_x,~\xi_2=\partial_y,~\xi_3=x\partial_y-y\partial_x$.
This metric can further be classified according to the additional
isometries it admits. For the sake of simplicity, we take the
coefficient of $dx^2$ as unity. The corresponding metric reduces
to
\begin{equation}
ds^2=e^{2\nu(t,x)}dt^2-dx^2-e^{2\lambda(t,x)}(dy^2+ dz^2).
\end{equation}
The energy-momentum tensor for a perfect fluid can be written as
\begin{equation}
T_{ab}=[\rho(t,x)+p(t,x)]u_{a}u_{b}- p(t,x)g_{ab},\quad
(a,b=0,1,2,3).
\end{equation}
where $\rho$ is the density, $p$ is the pressure and $u_{a}$ is
the four-velocity of the fluid element. In the co-moving
coordinate system, the four-velocity can be written as
$u_{a}=(e^{\nu(t,x)},0,0,0)$. The EFEs become
\begin{eqnarray}
\kappa\rho&=& e^{-2\nu}{\lambda_{t}}^{2}-3{\lambda_{x}}^{2}-2\lambda_{xx},\\
0&=&\lambda_{tx}-\lambda_{t}\nu_{x}+\lambda_{t}\lambda_{x},\\
\kappa p &=&{\lambda_{x}}^{2}+2\lambda_{x}\nu_{x}
-e^{-2\nu}(2\lambda_{tt}-2\lambda_{t}\nu_{t}+3{\lambda_{t}}^{2}),\\
\kappa p
&=&\nu_{xx}+{\nu_{x}}^{2}+\nu_{x}\lambda_{x}+{\lambda_{x}}^{2}
+\lambda_{xx}-e^{-2\nu}(\lambda_{tt}-\lambda_{t}\nu_{t}+{\lambda_{t}}^{2}).
\end{eqnarray}
The conservation of energy-momentum tensor, ${T^{ab}}_{;b}=0$,
provides the following two equations
\begin{equation}
\lambda_{t}=-\frac{\rho_{t}}{2(\rho+p)},
\end{equation}
and
\begin{equation}
\nu_{x}=\frac{p_{x}}{(\rho+p)}.
\end{equation}
The general form of a vector field $\xi$, for a plane symmetric
spacetime, can take the following form
\begin{equation}
\xi^{a}\frac{\partial}{\partial
x^a}=h_{1}(t,x)\frac{\partial}{\partial
t}+h_{2}(t,x)\frac{\partial}{\partial x},
\end{equation}
where $h_{1}$ and $h_{2}$ are arbitrary functions. When $\xi$ is
parallel to the fluid flow, $h_{2}=0$ and when $\xi$ is orthogonal
to the fluid flow $h_{1}=0$. When both $h_{1}$ and $h_{2}$ are
non-zero, $\xi$ is tilted to the fluid flow.

A KSS vector $\xi$ satisfies the following conditions
\begin{eqnarray}
\pounds_{\xi}h_{ab}&=& 2\delta h_{ab},\\
\pounds_{\xi}u_{a}&=& \alpha u_{a},
\end{eqnarray}
where $h_{ab}=g_{ab}-u_au_b$ is the projection tensor, $\alpha$
and $\delta$ are constants. The similarity transformation is
characterized by the scale independent ratio, $\alpha/\delta$.
This ratio is referred as the similarity index which yields the
following two cases according as:
\begin{eqnarray*}
1. \quad \delta\neq0,\\
2. \quad \delta=0.
\end{eqnarray*}
{\bf Case 1:} If $\delta\neq0$ it can be chosen as unity and the
KSS vector for the titled case can take the following form
\begin{equation}
\xi^{a}\frac{\partial}{\partial x^a}=(\alpha
t+\beta)\frac{\partial}{\partial t}+x\frac{\partial}{\partial x}.
\end{equation}
For this case, the similarity index, $\alpha/\delta$, further
yields the following three different possibilities.
\begin{eqnarray*}
(i)\quad\delta\neq0,\quad\alpha=1\quad
(\beta~can~be~taken~to~be~zero),\\
(ii)\quad\delta\neq0,\quad\alpha=0\quad
(\beta~can~be~taken~to~be~unity),\\
(iii)\quad\delta\neq0,\quad\alpha\neq0,1\quad
(\beta~can~be~taken~to~be~zero).
\end{eqnarray*}
The case 1(i) corresponds to the self-similarity of the {\it first
kind}. In this case $\xi$ is a homothetic vector and the
self-similar variable $\xi$ turns out to be $x/t$. For the second
case 1(ii), it is termed as the self-similarity of the {\it zeroth
kind} and the self-similar variable takes the following form
\begin{eqnarray*}
\xi=x e^{-t}.
\end{eqnarray*}
In the last case 1(iii), it is called the self-similarity of the
{\it second kind} and the self-similar variable becomes
\begin{eqnarray*}
\xi=\frac{x}{(\alpha t)^\frac{1}{\alpha}}.
\end{eqnarray*}
It turns out that for the case (1), when $\delta\neq0$, with the
self-similar variable $\xi$, the metric functions become
\begin{equation}
\nu=\nu(\xi),\quad e^{\lambda}=xe^{\lambda(\xi)}.
\end{equation}
The case (2), in which $\delta=0$ and $\alpha\neq0$ ($\alpha$ can
be unity and $\beta$ can be re-scaled to zero), the
self-similarity is known as the {\it infinite kind}. In this case,
the KSS vector $\xi$ turns out to be
\begin{equation}
\xi^{a}\frac{\partial}{\partial x^a}=t\frac{\partial}{\partial
t}+c\frac{\partial}{\partial x}
\end{equation}
and the self-similar variable will become
\begin{eqnarray*}
\xi=e^{\frac{x}{c}}/t,
\end{eqnarray*}
where $c$ is an arbitrary constant. The metric functions will be
of the following form
\begin{equation}
\nu=\nu(\xi),\quad\lambda=\lambda(\xi).
\end{equation}
Notice that for the plane symmetric spacetime, the self-similar
variable of the first, second and zeroth kinds turn out to be the
same as for the spherically and cylindrically symmetric spacetimes
with the exception that $r$ has been replaced by $x$ in the plane
symmetric metric. We note that for $\delta=0=\alpha$, the KSS
vector $\xi$ becomes KV.

If the KSS vector $\xi$ is parallel to the fluid flow, it follows
that
\begin{equation}
\xi^{a}\frac{\partial}{\partial x^a}= f(t)\frac{\partial}{\partial
t},
\end{equation}
where $f(t)$ is an arbitrary function. It is mentioned here that
the self-similar variable for spherically symmetric metric is $r$
whereas it turns out $r$ only for the infinite kind in the case of
cylindrically symmetric metric. In the remaining kinds, we obtain
contradictory results for the cylindrically symmetric spacetime.
For the plane symmetry, we obtain contradictory results in the
first, second and zeroth kinds while for the infinite kind the
self-similar variable turns out to be $x$. This implies that there
does not exist any solution when $\xi$ parallel to the fluid flow
in the first, second and zeroth kinds but there may be some
solution in the case of infinite kind.

When the KSS vector $\xi$ is orthogonal to the fluid flow, we
obtain
\begin{equation}
\xi^{a}\frac{\partial}{\partial x^a}= g(x)\frac{\partial}{\partial
x},
\end{equation}
where $g(x)$ is an arbitrary function and the self-similar
variable is $t$.

We assume the following two types of polytropic equation of states
(EOS). We denote the first equation of state by EOS(1) and is
given by
\begin{eqnarray*}
p=k\rho^{\gamma},
\end{eqnarray*}
where $k$ and $\gamma$ are constants. The other EOS can be written
as [17]
\begin{eqnarray*}
p=kn^{\gamma},\\
\rho=m_{b}n+\frac{p}{\gamma-1},
\end{eqnarray*}
where $m_{b}$ is a constant and corresponds to the baryon mass,
and $n(t,r)$ corresponds to baryon number density. This equation
is called second equation of state written as EOS(2). For EOS(1)
and EOS(2), we take $k\neq0$ and $\gamma\neq0,1$. The third
equation of state, denoted by EOS(3), is the following
\begin{eqnarray*}
p=k\rho.
\end{eqnarray*}
Here we assume that $-1\leq k \leq 1$ and $k \neq 0.$

For different values of $\gamma$, EOS(1) and EOS(2) have different
properties. Thermodynamical instability of the fluid is shown for
$\gamma<0.$ For $0<\gamma<1,$ both EOS(1) and EOS(2) are
approximated by a dust fluid in high density regime. For
$\gamma>1$, EOS(2) is approximated by EOS(3) with $k= \gamma-1$ in
high density regime. The cases $\gamma>2$ for EOS(2) and
$\gamma>1$ for EOS(2) shows that the dominant energy condition can
be violated in high density regime which is physically not
interesting [5].

\section{Tilted Perfect Fluid Case}

\subsection{Self-similarity of the first kind}

Firstly, we discuss the self-similarity of the first kind for the
tilted perfect fluid case. In this case, it follows from the EFEs
that the energy density $\rho$ and pressure $p$ must take the
following form
\begin{eqnarray}
\kappa\rho &=& \frac{1}{x^2}[\rho_1(\xi)+\frac{x^2}{t^2}\rho_2(\xi)],\\
\kappa p &=& \frac{1}{x^2}[p_1(\xi)+\frac{x^2}{t^2}p_2(\xi)],
\end{eqnarray}
where the self-similar variable is $\xi=x/t$. If the EFEs and the
equations of motion for the matter field are satisfied for
$O[(\frac{x}{t})^0]$ and $O[(\frac{x}{t})^2]$ terms separately, we
obtain a set of ordinary differential equations. Thus Eqs.(7)-(12)
reduce to the following
\begin{eqnarray}
-\dot{\rho_1} &=&2\dot{\lambda}(\rho_1+p_1),\\
\dot{\rho_2}+2\rho_2 &=&-2\dot{\lambda}(\rho_2+p_2),\\
\dot{p_1}-2p_1 &=& \dot{\nu}(\rho_1+p_1),\\
\dot{p_2} &=&\dot{\nu}(\rho_2+p_2) ,\\
-\rho_1 &=& 1+4\dot{\lambda}+3{\dot{\lambda}}^{2}+2\ddot{\lambda},\\
\rho_2 &=& {\dot{\lambda}}^{2}e^{-2\nu},\\
0&=&\ddot{\lambda}+{\dot{\lambda}}^2+\dot{\lambda}-\dot{\lambda}\dot{\nu},\\
p_1&=&1+2\dot{\lambda}+{\dot{\lambda}}^2+2\dot{\nu}+2\dot{\lambda}\dot{\nu},\\
-e^{2\nu}p_2 &=& 2\ddot{\lambda}+3{\dot{\lambda}}^2+2\dot{\lambda}
-2\dot{\lambda}\dot{\nu},\\
p_1 &=& \ddot{\lambda}+{\dot{\lambda}}^2+\dot{\lambda}
+\dot{\lambda}\dot{\nu}+\ddot{\nu}+{\dot{\nu}}^2,\\
-e^{2\nu}p_2&=&\ddot{\lambda}+{\dot{\lambda}}^2
+\dot{\lambda}-\dot{\lambda}\dot{\nu}.
\end{eqnarray}
where dot $(.)$ represents derivative with respect to $ln(\xi)$.
When we use Eq.(30) in Eq.(34), it turns out that $p_2=0$ which
together with Eqs.(30) and (32) yields that $\lambda$ is an
arbitrary constant. Solving the above equations simultaneously, we
get $\rho_1=-1$ from Eq.(28)and $\rho_2=0$ from Eq.(29). Finally,
we are left with three equations in two unknowns $p_1$ and $\nu$.
When we solve these equations simultaneously, we do not have such
values of $p_1$ and $\nu$ which satisfy these three equations.
Hence there is no solution in this case.

\subsection{Self-similarity of the second kind}

Here we discuss the self-similarity of the second kind for the
tilted perfect fluid case. In this case, it follows from the EFEs
that the energy density $\rho$ and pressure $p$ must take the
following form
\begin{eqnarray}
\kappa\rho &=& \frac{1}{x^2}[\rho_1(\xi)+\frac{x^2}{t^2}\rho_2(\xi)],\\
\kappa p &=& \frac{1}{x^2}[p_1(\xi)+\frac{x^2}{t^2}p_2(\xi)],
\end{eqnarray}
where the self-similar variable is $\xi=x/(\alpha
t)^\frac{1}{\alpha}$. If the EFEs and the equations of motion for
the matter field are satisfied for $O[(\frac{x}{t})^0]$ and
$O[(\frac{x}{t})^2]$ terms separately, we obtain a set of ordinary
differential equations. Thus Eqs.(7)-(12) take the following form
\begin{eqnarray}
\dot{\rho_1} &=&-2\dot{\lambda}(\rho_1+p_1),\\
\dot{\rho_2}+2\alpha\rho_2 &=&-2\dot{\lambda}(\rho_2+p_2),\\
\dot{p_1}-2p_1 &=& \dot{\nu}(\rho_1+p_1),\\
\dot{p_2} &=&\dot{\nu}(\rho_2+p_2) ,\\
0&=&\ddot{\lambda}+{\dot{\lambda}}^2+\dot{\lambda}
-\dot{\lambda}\dot{\nu},\\
-\rho_1 &=& 1+4\dot{\lambda}+3{\dot{\lambda}}^{2}+2\ddot{\lambda},\\
\alpha^2\rho_2 &=& {\dot{\lambda}}^{2}e^{-2\nu},\\
p_1&=&1+2\dot{\lambda}+{\dot{\lambda}}^2+2\dot{\nu}
+2\dot{\lambda}\dot{\nu},\\
-\alpha^2e^{2\nu}p_2 &=& 2\ddot{\lambda}+3{\dot{\lambda}}^2
+2\alpha\dot{\lambda}-2\dot{\lambda}\dot{\nu},\\
p_1&=&\ddot{\lambda}+{\dot{\lambda}}^2
+\dot{\lambda}+\dot{\lambda}\dot{\nu}+\ddot{\nu}+{\dot{\nu}}^2,\\
-\alpha^2e^{2\nu}p_2&=&\ddot{\lambda}+{\dot{\lambda}}^2
+\alpha\dot{\lambda}-\dot{\lambda}\dot{\nu}.
\end{eqnarray}
Now we solve this set of equations by using EOS (1)-(3).

\subsubsection{Equations of State (1) and (2)}

If a perfect fluid satisfies EOS(1) for $k\neq0$ and
$\gamma\neq0,1$, Eqs.(35) and (36) become
\begin{equation}
\alpha=\gamma,\quad p_1=0=\rho_2,\quad p_2= \frac{k}{(8\pi
G)^{(\gamma-1)}\gamma^2}\xi^{-2\gamma}{\rho_1}^\gamma,
\quad[Case~I]
\end{equation}
or
\begin{equation}
\alpha=\frac{1}{\gamma},\quad p_2=0=\rho_1,\quad
p_1=\frac{k}{(8\pi
G)^{(\gamma-1)}\gamma^{2\gamma}}\xi^{2}{\rho_2}^\gamma.\quad[Case~II]
\end{equation}
If a perfect fluid obeys EOS(2) for $k\neq0$ and $\gamma\neq0,1$,
we find from Eqs.(35) and (36) that
\begin{equation}
\alpha=\gamma,\quad p_1=0,\quad p_2=\frac{k}{{m_b}^{\gamma}(8\pi
G)^{(\gamma-1)}\gamma^2}\xi^{-2\gamma}{\rho_1}^\gamma=(\gamma-1)\rho_2,
\quad [Case~III]
\end{equation}
or
\begin{equation}
\alpha=\frac{1}{\gamma}, \quad p_2=0, \quad
p_1=\frac{k}{{m_b}^{\gamma}(8\pi
G)^{(\gamma-1)}\gamma^{2\gamma}}\xi^{2}{\rho_2}^\gamma=(\gamma-1)\rho_1.
\quad [Case~IV]
\end{equation}
In all the above cases, we obtain contradiction to the basic
equations and we can conclude that there is no solution in the
above mentioned cases.

\subsubsection{Equation of State (3)}

If a perfect fluid satisfies EOS(3), Eqs.(35) and (36) yield that
\begin{equation}
p_1=k\rho_1, \quad p_2=k\rho_2. \quad [Case~V]
\end{equation}
When $k=-1$, we have a contradiction in the basic Eqs.(37)-(47).
For $k\neq-1$, we assume that $\rho_1\neq0$ and $\rho_2\neq0$. In
this case, using Eqs.(37) and (38), it follows that
$-2\alpha\rho_1\rho_2-\dot{\rho_2}{\rho_1}+\dot{\rho_1}{\rho_2}=0,$
and from Eqs.(39) and (40),  we obtain
$-2\rho_1\rho_2-\dot{\rho_2}{\rho_1}+\dot{\rho_1}{\rho_2}=0.$
These two expressions imply that $\rho_1\rho_2=0$ as
$\alpha\neq1$. For the case when $\rho_1=0=p_1$ and $\rho_2\neq0$,
we have a contradiction. The case when $\rho_2=0=p_2$ and
$\rho_1\neq0$, Eq.(43) implies that $\dot{\lambda}=0$, i.e.,
$\lambda=constant$. Now Eq.(42) shows that $\rho_1=-1,$ and EOS(3)
implies that $p_1=-k$. Using these information in the basic
equations, we get two values of $\dot{\nu}$ from Eq.(39) and
Eq.(44). Comparing these values we get $k=1$ and then it follows
that $\nu=\ln(\frac{b_0}{\xi})$ from both equations. On
substituting the values of $k$ and $\nu$ in Eq.(46), we reach at a
contradiction. Hence, there is no self-similar solution in this
case.

\subsection{Self-similarity of the zeroth kind}

This section is devoted for the discussion of the self-similar
solutions of the zeroth kind. In this case, the EFEs indicate that
the quantities $\rho$ and $p$ must be of the form
\begin{eqnarray}
\kappa\rho &=& \frac{1}{x^2}[\rho_1(\xi)+x^2\rho_2(\xi)],\\
\kappa p &=& \frac{1}{x^2}[p_1(\xi)+x^2p_2(\xi)],
\end{eqnarray}
where the self-similar variable is $\xi=\frac{x}{e^{-t}}$. If it
is assumed that the EFEs and the equations of motion for the
matter field are satisfied for $O[(x)^0]$ and $O[(x)^2]$ terms
separately, we obtain the following set of ordinary differential
equations.
\begin{eqnarray}
\dot{\rho_1} &=&-2\dot{\lambda}(\rho_1+p_1),\\
\dot{\rho_2} &=&-2\dot{\lambda}(\rho_2+p_2),\\
\dot{p_1}-2p_1 &=& \dot{\nu}(\rho_1+p_1),\\
\dot{p_2} &=&\dot{\nu}(\rho_2+p_2) ,\\
0&=&\ddot{\lambda}+{\dot{\lambda}}^2+\dot{\lambda}-\dot{\lambda}\dot{\nu},\\
-\rho_1 &=& 1+4\dot{\lambda}+3{\dot{\lambda}}^{2}+2\ddot{\lambda},\\
\rho_2 &=& {\dot{\lambda}}^{2}e^{-2\nu},\\
p_1&=&1+2\dot{\lambda}+{\dot{\lambda}}^2+2\dot{\nu}+2\dot{\lambda}\dot{\nu},\\
-e^{2\nu}p_2&=&2\ddot{\lambda}+3{\dot{\lambda}}^2-2\dot{\lambda}\dot{\nu},\\
p_1&=&\ddot{\lambda}+{\dot{\lambda}}^2+\dot{\lambda}+\dot{\lambda}\dot{\nu}
+\ddot{\nu}+{\dot{\nu}}^2,\\
-e^{2\nu}p_2&=&\ddot{\lambda}+{\dot{\lambda}}^2-\dot{\lambda}\dot{\nu}.
\end{eqnarray}

\subsubsection{EOS(1) and EOS(2)}

These two EOS for the zeroth kind give a contradiction and hence
there does not exist any solution.

\subsubsection{EOS(3)}

When we take a perfect fluid satisfying EOS(3), it follows from
Eqs.(53) and (54) that
\begin{equation}
p_1=k\rho_1,\quad  p_2=k\rho_2.
\end{equation}
We can proceed in a similar way as in the case of self-similarity
of the second kind with EOS(3). and met contradiction in each
case.

\subsection{Self-similarity of the infinite kind}

In this section we discuss the self-similar solution of the
infinite kind. For this case, the EFEs imply that the quantities
$\rho$ and $p$ must be of the form
\begin{eqnarray}
\kappa\rho &=& \frac{1}{t^2}\rho_1(\xi)+\frac{1}{c^2}\rho_2(\xi),\\
\kappa p &=& \frac{1}{t^2}p_1(\xi)+\frac{1}{c^2}p_2(\xi),
\end{eqnarray}
where $\xi=\frac{e^\frac{x}{c}}{t}.$  Now if we require that the
EFEs and the equations of motion for the matter field are
satisfied for $O[(t)^0]$ and $O[(t)^{-2}]$ terms separately, we
obtain a set of ordinary differential equations. For a perfect
fluid, Eqs.(7)-(12) takes the following form
\begin{eqnarray}
\dot{\rho_1}+2\rho_1 &=&-2\dot{\lambda}(\rho_1+p_1),\\
\dot{\rho_2} &=&-2\dot{\lambda}(\rho_2+p_2),\\
\dot{p_1} &=& \dot{\nu}(\rho_1+p_1),\\
\dot{p_2} &=&\dot{\nu}(\rho_2+p_2),\\
0&=&\ddot{\lambda}+{\dot{\lambda}}^2-\dot{\lambda}\dot{\nu},\\
\rho_1 &=& {\dot{\lambda}}^{2}e^{-2\nu},\\
\rho_2 &=& -3{\dot{\lambda}}^{2}-2\ddot{\lambda} ,\\
-e^{2\nu}p_1&=& 2\ddot{\lambda}+3{\dot{\lambda}}^2
+2\dot{\lambda}-2\dot{\lambda}\dot{\nu},\\
p_2&=&{\dot{\lambda}}^2+2\dot{\lambda}\dot{\nu},\\
-e^{2\nu}p_1&=&\ddot{\lambda}+{\dot{\lambda}}^2
+\dot{\lambda}-\dot{\lambda}\dot{\nu},\\
p_2&=&\ddot{\lambda}+{\dot{\lambda}}^2+\dot{\lambda}\dot{\nu}
+\ddot{\nu}+{\dot{\nu}}^2,
\end{eqnarray}
respectively.

\subsubsection{EOS(1) and EOS(2)}

When a perfect fluid satisfies EOS(1), it can be seen from Eq.(67)
and Eq.(68) that
\begin{eqnarray}
p_1=0=\rho_1,\quad p_2= k(8\pi
G)^{(1-\gamma)}{\rho_2}^\gamma.\quad [Case~I]
\end{eqnarray}
For the condition given by EOS(2), it turns out that
\begin{eqnarray}
p_1=0=\rho_1,\quad p_2=\frac{k}{{m_b}^{\gamma}(8\pi
G)^{(\gamma-1)}}{(\rho_2-\frac{p_2}{( \gamma-1)})}^\gamma.\quad
[Case~II]
\end{eqnarray}
In both cases, Eq.(74) shows that $\lambda=constant$ and then
Eq.(75) gives $\rho_2=0.$ Now EOS(1) and EOS(2) show that $p_2=0.$
We are left with Eq.(79) which gives $\ddot{\nu}+{\dot{\nu}}^2=0$
and consequently $\nu=\ln[c_1(\ln\xi-c_2)]$. Finally, we have the
following vacuum solution
\begin{eqnarray}
\nu=\ln[ax-b\ln t-c],\quad \lambda=contant,\nonumber\\
\rho_1=0=p_1,\quad \rho_2=0=p_2.
\end{eqnarray}

\subsubsection{EOS(3)}

It follows from Eqs.(67) and (68) that this equation of state
gives
\begin{equation}
p_1=k\rho_1,\quad  p_2=k\rho_2.\quad [Case~III]
\end{equation}
We consider the following two possibilities when $k=-1$ and
$k\neq-1$. In the first case, we have
\begin{equation}
p_1+\rho_1=0,\quad p_2+\rho_2=0.
\end{equation}
If we make use of Eqs.(69)-(72), we obtain $\rho_1=0=p_1$ and
$\rho_2=-p_2=constant$. Then it follows from Eq.(74) that
$\lambda=constant$ and Eq.(75) gives $\rho_2=0$. Also, from
EOS(3), we can say that $p_2=0$. This turns out exactly the same
solution as in the case of EOS(1) and EOS(2).

In the second case, i.e., $k\neq-1$ when $\rho_1\neq0$ and
$\rho_2\neq0$, solving Eqs.(69)-(72) simultaneously, we have
$\rho_1\rho_2=0$. If we consider $\rho_1=0=p_1$, or
$\rho_2=0=p_2$, we again have the same results as in EOS(1) and
EOS(2).

\section{Tilted Dust Case}

\subsection{Self-similarity of the first kind}

If we set $p_1=0=p_2$ in the basic Eqs.(24)-(34) for the tilted
perfect fluid case with self-similarity of the first kind,
Eqs.(30) and (32) immediately gives $\lambda=0$. From the rest of
the equations, Eq.(26) and Eq.(27) give rise to the two cases
either $\nu=constant$ or $\rho_1=0=\rho_2$. The first possibility
contradicts Eq.(31) and the second possibility contradicts
Eq.(28). Hence there does not exist any solution for this case.

\subsection{Self-similarity of the second kind}

When we take $p_1=0=p_2$ in Eqs.(37)-(47) for the tilted perfect
fluid case with self-similarity of the second kind, Eqs.(39) and
(40) immediately gives either $\nu=constant$ or $\rho_1=0=\rho_2$.
This again leads to the contradiction.

\subsection{Self-similarity of zeroth kind}

In this case, we also have contradiction and consequently there is
no solution.

\subsection{Self-similarity of infinite kind}

In this case we take $p_1=0=p_2$ in Eqs.(69)-(79) for the tilted
perfect fluid case with self-similarity of the infinite kind.
Eqs.(71) and (72) imply that either $\nu=constant$ or
$\rho_1=0=\rho_2$. In the first case, Eq.(77) directly gives
$\lambda=constant$ and Eqs.(74) and (75) show that
$\rho_1=0=\rho_2$ and hence we have a Minkowski spacetime. For the
second case, Eq.(74) implies that $\lambda=constant$ and we are
left with Eq.(79) only which gives $\ddot{\nu}+{\dot{\nu}}^2=0$.
This yields exactly the similar result as for the perfect fluid
with self-similarity of the infinite kind using EOS(1) and EOS(2).

\section{Orthogonal Perfect Fluid Case}

\subsection{Self-similarity of the first kind}

Here we discuss self-similar solutions for the orthogonal perfect
fluid case. Firstly we consider the self-similarity of the first
kind. In this case, the self-similar variable is $\xi=t$ and the
plane symmetric spacetime becomes
\begin{equation}
ds^2=x^2e^{2\nu(t)}dt^2- dx^2-x^2e^{2\lambda(t)}(dy^2+dz^2).
\end{equation}
EFEs and the equations of motion for the matter field gives the
following set of equations
\begin{eqnarray}
e^{2\nu}(1+\rho)&=& {\lambda'}^2,\\
e^{2\nu}(3-p)&=&3{\lambda'}^2+2\lambda''-2\lambda'\nu',\\
e^{2\nu}(1-p)&=&\lambda''+{\lambda'}^2-\lambda'\nu',\\
2\lambda'(\rho+p)&=&-\rho'_1,\\
\rho+3p=0,
\end{eqnarray}
where prime represents derivative with respect to $\xi=t$. Eq.(90)
gives us an equation of state for this system of equations. Using
this EOS in Eq.(89), we can get the value of $\lambda'$ in terms
of pressure as $\lambda'=-\frac{3p'}{4p}.$ Also, using this value
of $\lambda'$ in Eq.(86), we can have the value of $e^{2\nu}$ in
terms of pressure as
$e^{2\nu}=\frac{9{p'}_1^2}{16p_1^2(1-3p_1)^2}$. Solving Eqs.(87)
and (88) simultaneously, we get contradiction to the value of
$e^{2\nu}$. This implies that no solution exist for this case.

\subsection{Self-similarity of the second kind}

Now we consider the self-similarity of the second kind. For this
case, the self-similar variable is given by $t.$ The plane
symmetric spacetime takes the form
\begin{equation}
ds^2=x^{2\alpha}e^{2\nu(t)}dt^2-
dx^2-x^2e^{2\lambda(t)}(dy^2+dz^2),
\end{equation}
EFEs imply that the quantities $\rho$ and $p$ must be of the form
\begin{eqnarray}
\kappa\rho &=&x^{-2}\rho_1(\xi)+x^{-2\alpha}\rho_2(\xi),\\
\kappa p &=& x^{-2}p_1(\xi)+x^{-2\alpha}p_2(\xi),
\end{eqnarray}
where $\xi=t$. We note that the solution is always singular at
$x=0$ which corresponds to the physical center. When the EFEs and
the equations of motion for the matter field are satisfied for
$O[(x)^0]$ and $O[(x)^{-2-2\alpha}]$ terms separately, we obtain a
set of ordinary differential equations. These are given as
\begin{eqnarray}
\rho_1 &=& -1,\\
\rho_2 &=& e^{-2\nu}{\lambda'}^2,\\
0 &=& (1-\alpha)\lambda',\\
p_1 &=& 1+2\alpha,\\
e^{2\nu}p_2&=&-2\lambda''+2\lambda'\nu'-3{\lambda'}^2,\\
p_1&=&\alpha^2,\\
-e^{2\nu}p_2&=&\lambda''-\lambda'\nu'+{\lambda'}^2,\\
2\lambda'(\rho_1+p_1)&=& -\rho'_1,\\
2\lambda'(\rho_2+p_2)&=& -\rho'_2,\\
\alpha(\rho_1+p_1)&=&-2p_1,\\
\rho_2+3p_2 &=& 0.
\end{eqnarray}
Eq.(104) gives us an EOS for this system. Since $p_1=0$
contradicts Eq.(99) and $\rho_1=0$ contradicts Eq.(94), a vacuum
spacetime is not compatible with this case. Eq.(96) gives us
$\lambda=constant$ and also Eq.(95) shows that $\rho_2=0$. From
Eq.(104), we have $p_2=0$ and also Eqs.(97) and (99) yield
$\alpha=1\pm\sqrt{2}$ and consequently $p_1=3\pm2\sqrt{2}$. These
values contradict our basic equations and hence we have no
solution in this case as well.

\subsection{Self-similarity of the zeroth kind}

The self-similar variable for this kind is also $t$ and the metric
for the plane symmetry becomes
\begin{equation}
ds^2=e^{2\nu(t)}dt^2- dx^2-x^2e^{2\lambda(t)}(dy^2+dz^2).
\end{equation}
In the case of self-similarity of the zeroth kind, the basic
equations for perfect fluid gives us a contradiction and hence we
have no solution in this case.

\subsection{Self-similarity of the infinite kind}

For the self-similarity of the infinite kind, self-similar
variable is $\xi=t$. The metric for this kind takes the following
form
\begin{equation}
ds^2=e^{2\nu(t)}dt^2- dx^2-e^{2\lambda(t)}(dy^2+dz^2).
\end{equation}
A set of ordinary differential equations in terms of $\xi$ is
obtained from EFEs and the equations of motion for the matter
field
\begin{eqnarray}
2\lambda'(\rho+p) &=& -\rho',\\
\rho &=&{\lambda'}^2e^{-2\nu},\\
-e^{2\nu}p&=&2\lambda''+3{\lambda'}^2-2\lambda'\nu',\\
-e^{2\nu}p&=&\lambda''+{\lambda'}^2-\lambda'\nu'.
\end{eqnarray}
We use EOS(3) to solve the set of  Eqs.(107)-(110) as this is the
only compatible equation of state for this kind.

\subsubsection{EOS(3)}

This equation of state is given by
\begin{equation}
p=k\rho.
\end{equation}
Using EOS(3) in Eqs.(109) and (110), we have
\begin{eqnarray}
-e^{2\nu}k\rho&=&2\lambda''+3{\lambda'}^2-2\lambda'\nu',\\
-e^{2\nu}k\rho&=&\lambda''+{\lambda'}^2-\lambda'\nu'.
\end{eqnarray}
Now putting the value of ${\lambda'}^2$ in the above two equations
from Eq.(108) and using Eqs.(112) and Eq.(113), we have
\begin{equation}
-e^{2\nu}\rho(k+1)=0.
\end{equation}
This gives two possibilities either $\rho=0$ or $k=-1$. For the
first possibility we immediately get from EOS(3) that $p=0$ and
from Eq.(108) $\lambda=constant$. Thus we obtain the following
vacuum solution
\begin{eqnarray}
\nu=\nu(t),\quad \lambda=a_0,\quad \rho=0=p.
\end{eqnarray}
For the second possibility, EOS(3) becomes $\rho+p=0$. Solving
Eqs.(107)-(110) simultaneously, we obtain the same solution as for
the first possibility.

\section{Orthogonal Dust case}

\subsection{Self-similarity of the first kind}

In this case we substitute $p=0$ in the basic equations for the
orthogonal perfect fluid case with the self-similarity of first
kind. Eq.(90) immediately shows that the resulting spacetime must
be vacuum. Eq.(86) gives $e^{2\nu}={\lambda'}^2.$ Also, on solving
Eqs.(87) and (88) we reach at $e^{2\nu}={\lambda'}^2.$ For this
case, we have the following solution
\begin{eqnarray}
\lambda=\int e^{\nu}dt,\quad \rho=0=p.
\end{eqnarray}
For $\nu=0$, this leads to the following solution
\begin{eqnarray}
\lambda=t,\quad \nu=0, \quad \rho=0=p.
\end{eqnarray}
\subsection{Self-similarity of the second  and zeroth kinds}

For the self-similarity of second and zeroth kinds we arrive at
the contradiction and hence there is no solution.

\subsection{Self-similarity of the infinite kind}

We set $p=0$ in the basic equations for the orthogonal perfect
fluid case for the self-similarity of infinite kind. Eqs.(109) and
(110)give $\lambda=constant$. Now Eq.(108) ensures us that the
resulting spacetime must be vacuum. Hence we have the same
solution as in the case of orthogonal perfect fluid case of self
similarity of infinite kind. This is given by
\begin{eqnarray}
\nu=\nu(t),\quad \lambda=a_0,\quad \rho=0=p.
\end{eqnarray}

\section{Parallel Perfect Fluid Case}

Since we do not have self-similar variable for the first, second
and zeroth kinds in the parallel perfect fluid case hence there
does not exist any solution for these kinds.

\subsection{Self-similarity of the infinite kind}

For the self-similarity of the infinite kind, self-similar
variable is $\xi=x$. The metric for the plane symmetry of the
infinite kind reduces to
\begin{equation}
ds^2=e^{2\nu(x)}dt^2- dx^2-e^{2\lambda(x)}(dy^2+dz^2).
\end{equation}
A set of ordinary differential equations in terms of $\xi$ is
obtained from EFEs and the equations of motion for the matter
field are
\begin{eqnarray}
-\rho&=& 3{\lambda'}^2+2\lambda'',\\
p &=&{\lambda'}^2+2\lambda'\nu',\\
p&=&\lambda''+{\lambda'}^2+\lambda'\nu'+\nu''+{\nu'}^2,\\
p'&=&\nu'(\rho+p).
\end{eqnarray}
Here prime $(')$ represents derivative with respect to $\xi=x$. We
use EOS(3) to solve the set of  Eqs.(120)-(123).

\subsubsection{EOS(3)}

This equation of state is given by
\begin{equation}
p=k\rho.
\end{equation}
Clearly Eq.(123) gives the value of $\nu'$ in terms of $\rho$ as
$\nu'=\frac{k\rho'}{(k+1)\rho}$. Now we assume, for the sake of
simplicity, $\rho$ as linear function of $\xi$. When we use this
assumption in Eq.(120), we obtain the value of $\lambda$ but
Eq.(121) gives a contradiction. Hence there does not exist any
self-similar solution for the parallel perfect fluid case.

\section{Parallel Dust case}

Again we do not have any self-similar variable for the first,
second and zeroth kinds in this case. Consequently, there does not
exist any solution for these kinds.

\subsection{Self-similarity of the infinite kind}

When we set $p=0$ in the basic equations for the parallel perfect
fluid case with self-similarity of the infinite kind, Eq.(123)
immediately shows that either $\nu=constant$ or $\rho=0$. In the
first case, Eqs.(120) and (121) show that the resulting spacetime
is \textit{Minkowski}. In the second case, where $\rho=0$,
Eq.(121) shows that either $\lambda'=0$ or $\lambda'=-2\nu'$. When
$\lambda'=0$, Eq.(122) gives the following vacuum solution
\begin{eqnarray}
\nu=ln(c_1(\xi-c_2)),\quad \lambda=constant=c_0,\quad \rho=0=p.
\end{eqnarray}
For the case $\lambda'=-2\nu'$, Eqs.(120) and (122) gives the
following vacuum solution
\begin{eqnarray}
\nu=-\frac{1}{3}ln(3\xi-2c_1),\quad
\lambda=ln(c_2(3\xi-2c_1)^\frac{2}{3}),\quad \rho=0=p.
\end{eqnarray}

\section{Conclusion}

We have classified KSS perfect fluid and dust solutions for the
cases when KSS vector is tilted, orthogonal and parallel to the
fluid flow by using EOS(1), EOS(2) and EOS(3). In most of the
cases, we solve the governing equations to get solutions but few
exceptions are there. We obtain total of six independent plane
symmetric self-similar solutions. The parallel case gives a
contradiction for the first, second, and zeroth kinds hence there
is no self-similar solution in these cases.

For the tilted perfect fluid case with self-similarity of the
first kind, we have contradictory results and hence no solution in
this case. For the self-similarity of the second and zeroth kinds
with EOS(1) and EOS(2), we again reach at a contradiction. For
self-similarity of the second and zeroth kinds with EOS(3), we
have a contradiction in all cases. For self-similarity of the
infinite kind, we find that the spacetime must be vacuum for
EOS(1) and EOS(2). It also turns out a vacuum solution for EOS(3)
when $k=-1$. For the case when $k\neq-1$ and either $\rho_1=0$ or
$\rho_2=0$ , we again have vacuum spacetime.

In the tilted dust case with self-similarity of the infinite kind,
we obtain two possibilities. One possibility gives a vacuum
solution and another case yields Minkowski spacetime. There is no
solution in any other kind.

For the orthogonal perfect fluid case with self-similarity of the
first kind, we have no solution. In the orthogonal perfect fluid
case with self-similarity of the second kind and zeroth kind, we
have a contradiction and hence there is no solution in these
cases. For the infinite kind with EOS(3) we have a vacuum solution
with arbitrary $\nu$ and constant $\lambda$.

In orthogonal dust case with self-similarity of the first kind, we
have a vacuum spacetime, where $\nu$ and $\lambda$ are related
with each other. For $\nu=0$,  $\lambda$ simply becomes $t$. We
have contradictory results for the self-similarity of the second
and zeroth kinds hence there is no solution in this case. We
obtain the same solution in the infinite kind as for the
orthogonal perfect fluid of the infinite kind. In the parallel
perfect fluid case, there does not exist any self-similar
solution. However, we obtain three different solutions for the
parallel dust case with self-similarity of the infinite kind.

We would like to mention here that this paper has been focussed on
a classification of plane symmetric kinematic self-similar
solutions under certain restrictions. A classification for the
most general plane symmetric kinematic self-similar solutions is
under investigation [23] and will appear somewhere else.

The results can be summarized in the form of tables given below:

\newpage

{\bf {\small Table 1.}} {\small Perfect fluid kinematic
self-similar solutions}.

\vspace{0.5cm}

\begin{center}
\begin{tabular}{|l|l|}
\hline {\bf Self-similarity} & {\bf Solution}
\\ \hline First kind (tilted) & None
\\ \hline First kind (orthogonal) & None
\\ \hline First kind (parallel) & None
\\ \hline Second kind (tilted)(EOS(1)) & None
\\ \hline Second kind (tilted)(EOS(2)) & None
\\ \hline Second kind (tilted)(EOS(3)) & None
\\ \hline Second kind (orthogonal) & None
\\ \hline Second kind (parallel) & None
\\ \hline Zeroth kind (tilted)(EOS(1)) & None
\\ \hline Zeroth kind (tilted)(EOS(2)) & None
\\ \hline Zeroth kind (tilted)(EOS(3)) & None
\\ \hline Zeroth kind (orthogonal) & None
\\ \hline Zeroth kind (parallel) & None
\\ \hline Infinite kind (tilted)(EOS(1)) & vacuum solution given by Eq.(82)
\\ \hline Infinite kind (tilted)(EOS(2)) & vacuum solution given by Eq.(82)
\\ \hline Infinite kind (tilted)(EOS(3)) & vacuum solution given by Eq.(82)
\\ \hline Infinite kind (orthogonal)(EOS(3)) & vacuum solution given by Eq.(115)
\\ \hline Infinite kind (parallel)(EOS(3)) & None
\\ \hline
\end{tabular}
\end{center}

\newpage

{\bf {\small Table 2.}} {\small Dust kinematic self-similar
solutions }.

\vspace{0.5cm}

\begin{center}
\begin{tabular}{|l|l|}
\hline {\bf Self-similarity} & {\bf Solution}
\\ \hline First kind (tilted) & None
\\ \hline First kind (orthogonal) & solution given by Eq.(116)
\\ \hline First kind (parallel) & None
\\ \hline Second kind (tilted) & None
\\ \hline Second kind (orthogonal) & None
\\ \hline Second kind (parallel) & None
\\ \hline Zeroth kind (titled) & None
\\ \hline Zeroth kind (orthogonal) & None
\\ \hline Zeroth kind (parallel) & None
\\ \hline Infinite kind (tilted) (case1)& Minkowski
\\ \hline Infinite kind (tilted) (case2)& vacuum solution given by Eq.(82)
\\ \hline Infinite kind (orthogonal) & vacuum solution given by Eq.(115)
\\ \hline Infinite kind (parallel)(case1) & Minkowski
\\ \hline Infinite kind (parallel)(case2) & vacuum solution given by Eq.(125)
\\ \hline Infinite kind (parallel)(case3) & vacuum solution given by Eq.(126)
\\ \hline
\end{tabular}
\end{center}

\vspace{1cm}

\begin{description}
\item  {\bf Acknowledgment}
\end{description}

One of us (SA) acknowledge the enabling role of the Higher
Education Commission Islamabad, Pakistan and appreciate its
financial support through {\it Merit Scholarship Scheme for Ph.D.
Studies in Science and Technology (200 Scholarships)}.

\newpage

{\bf \large References}

\begin{description}

\item{[1]} Noether, E. Nachr, Akad. Wiss. Gottingen, II, Math. Phys. {\bf
K12}(1918)235;\\
Davis, W.R. and Katzin, G.H. Am. J. Phys. {\bf 30}(1962)750;\\
Petrov, A.Z.: {\it Einstein Spaces} (Pergamon, Oxford University
Press, 1969);\\
Hojman, L. Nunez, L. Patino, A. and Rago, H.: J. Math. Phys. {\bf
27} (1986)281.

\item{[2]} Carr, B.J.: Phys. Rev. {\bf D62}(2000)044022.

\item{[3]} Carr, B.J. and Coley, A.A.: Phys. Rev. {\bf D62}(2000)044023.

\item{[4]} Maeda, H., Harada, T., Iguchi, H. and Okuyama, N.: Phys. Rev.
           {\bf D66}(2002)027501.

\item{[5]} Maeda, H., Harada, T., Iguchi, H. and Okuyama, N.: Prog. Theor. Phys.
           {\bf108}(2002)819-851.

\item{[6]} Maeda, H., Harada, T., Iguchi, H. and Okuyama, N.: Prog. Theor. Phys.
           {\bf110}(2003)25-63.

\item{[7]} Sharif, M.: J. Math. Phys. {\bf44}(2003)5141;
ibid {\bf45}(2004)1518; ibid {\bf45}(2004)1532.

\item{[8]} Sharif, M. and Aziz, Sehar: Int. J. Mod. Phys. {\bf D14}(2005)1527;
           \\{\it Kinematic Self-Similar Solutions: Proceedings of the 11th
           Regional Conference on Mathematical
           Physics and IPM Spring Conference} Tehran-Iran, May, 3-6, 2004,
           eds. Rahvar, S., Sadooghi, N. and Shojai, F. (World Scientific,
           2005)111.

\item{[9]} Penston, M.V.: Mon. Not. R. Astr. Soc.
           {\bf144}(1969)425.\\
           Larson, R.B.: Mon. Not. R. Astr. Soc.
           {\bf145}(1969)271.\\
           Shu, F.H.:  Astrophys. J. {\bf214}(1977)488.\\
           Hunter, C.: Astrophys. J. {\bf218}(1977)834.

\item{[10]} Cahill, M.E. and Taub, A.H.: Commun. Math. Phys.
           {\bf21}(1971)1.

\item{[11]} Carter, B. and Henriksen, R.N.: Annales De Physique
           {\bf14}(1989)47.

\item{[12]} Carter, B. and Henriksen, R.N.: J. Math. Phys.
           {\bf32}(1991)2580.

\item{[13]} Coley, A.A.: Class. Quant. Grav. {\bf14}(1997)87.

\item{[14]} McIntosh, C.B.G.: Gen. Relat. Gravit. {\bf7}(1975)199.

\item{[15]} Benoit, P.M. and Coley, A.A.: Class. Quant. Grav.
           {\bf15}(1998)2397.

\item{[16]} Carr, B.J., Coley, A.A., Golaith, M., Nilsson, U.S. and Uggla, C.:
           Class. Quant. Grav. {\bf18}(2001)303.

\item{[17]} Carr, B.J., Coley, A.A., Golaith, M., Nilsson, U.S. and Uggla, C.:
            Phys. Rev. {\bf D61}(2000)081502.

\item{[18]} Coley, A.A. and Golaith, M.: Class. Quant. Grav. {\bf17}(2000)2557.

\item{[19]} Sharif, M. and Aziz, Sehar: Int. J. Mod. Phys. {\bf D14}(2005)73.

\item{[20]} Sharif, M. and Aziz, Sehar: Int. J. Mod. Phys. {\bf A20}(2005)7579.

\item{[21]} Sharif, M. and Aziz, Sehar: J. Korean Physical Society {\bf 47}(2005)757.

\item{[22]} Stephani, H., Kramer, D., Maccallum, M., Hoenselaers, C. and Herlt,E.
            \textit{Exact Solutions of Einstein's Field Equations}
            (Cambridge University Press, 2003).

\item{[23]} Sharif, M. and Aziz, Sehar: work in progress.

\end{description}

\end{document}